\documentstyle[eqsecnum,prb,aps]{revtex}  

\begin{document}                
\baselineskip 0.5cm
%

\title{Transport and magnetic anomalies due to A-site ionic-size mismatch 
        in La$_{0.5}$Ca$_{0.5-x}$Ba$_x$MnO$_3$}
\author{R. Mallik, E.S. Reddy, P.L. Paulose, Subham Majumdar and E.V.
Sampathkumaran$^*$}
\address{\it Tata Institute of Fundamental Research, Homi Bhabha Road,
Colaba, Mumbai - 400 005, INDIA}
\maketitle

\begin{abstract}
We present the results of electrical resistivity ($\rho$),
magnetoresistance (MR) and dc and ac susceptibility ($\chi$) measurements on
polycrystalline samples of the type, La$_{0.5}$Ca$_{0.5-x}$Ba$_x$MnO$_3$,
synthesized under identical heat-treatment conditions.  The substitution
of larger Ba ions for Ca  results in a {\it non-monotonic} variation of
Curie temperature (T$_C$) as the system evolves from a charge-ordered
insulating state for $x$= 0.0 to a ferromagnetic metallic state for $x$=
0.5. An intermediate composition, x= 0.1, interestingly exhibits
ferromagnetic, insulating behavior with thermal hysteresis in ac $\chi$
around Curie temperature (T$_C$= 120 K).  The $x$= 0.2 and 0.3 compounds
exhibit semi-conducting-like behavior as the temperature is lowered below
300 K, with a broad peak in $\rho$ around 80-100 K; these compositions
exhibit a weak increase in $\rho$ as the temperature is lowered  below 30
K, indicative of electron localisation effects; these compositions also
undergo ferromagnetic transition below about 200 and 235 K respectively,
though these are non-hysteretic;  above all, for these compositions, MR is
large and conveniently measurable over the entire temperature range of
measurement below T$_C$ and this experimental finding  may be of interest
from the applications point of view.  We infer that the A-site ionic-size
mismatch plays  a crucial role in deciding these properties.
\end{abstract}

\vskip 1cm
\hskip 2cm PACS numbers:   Principal : 71.30.+h. 
Additional: 71.27.+a;   75.25.+z

\twocolumn
\vskip 1cm
The observation of large magnetoresistance (MR)  in the vicinity of  Curie
temperature (T$_C$) in hole-doped LaMnO$_3$ systems, both in thin film as
well as polycrystals,\cite{1,2} opened up a new direction of research,
viz., giant magnetoresistance (see, for instance, Refs. 3-4). From the
applications point of view, it is desirable\cite{5} to identify systems
exhibiting conveniently measurable electrical resistivity ($\rho$) and a
large MR over a wide temperature range.  In this sense, the observation of
such an effect in polycrystalline thin films as well as in bulk form (but
not in epitaxial thin films and bulk single crystals) of
La$_{2/3}$(Ba,Ca)$_{1/3}$MnO$_3$ is of considerable interest.\cite{5,6,7}
Apparently, the presence of strains across the grain boundaries is an
important deciding factor for such a behavior.\cite {6} In this respect,
we considered it worthwhile to carry out investigations on the series,
La$_{0.5}$Ca$_{0.5-x}$Ba$_x$MnO$_3$, as there is a large lattice strain
induced by size-mismatch between Ca and Ba.  Our results reported here
establish  large magnetoresistance for x= 0.2 and 0.3 compositions over a
wide temperature range as the temperature is lowered below T$_C$;  the
magnitude of MR is precisely measurable down to 4.2 K considering that
these compositions are not insulators at low temperatures in zero field.

There is also another motivation for the investigation of this Ba
substituted series. It is known that the x= 0.0 oxide exhibits  a charge
ordered (CO) antiferromagnetic (AF) insulating state below 150 K and the
ferromagnetism sets in below 220 K (see the articles cited in Ref. 8). The
substitution of a bigger divalent cation, Sr$^{2+}$ for Ca$^{2+}$,  in the
above series enhances the strength of double exchange (DE) interaction;
this results in a gradual transformation of the insulating
antiferromagnetic CO state into a DE ferromagnetic metal as the Sr
concentration is increased resulting in a monotonic increase of
T$_C$.\cite{8} While  the magnetic and transport behavior sensitively
depend on the size of A-site average ionic radius in general in these
perovskites,\cite{9} there are few reports emphasizing the role of
size-mismatch of the ions at the A-site on the properties.\cite{10,11} It
is therefore of interest to explore how the properties are modified by a
higher degree of lattice strain, which can be induced by substituting Ba
for Ca considering  a large variation in the average A-site radius from
1.198\AA   to 1.343 \AA   for x= 0.0 and 0.5 respectively (while
substituting Ba for Ca).

At this juncture, we would like to remark that there is very little work
in the literature on the influence of Ba substitution for Ca in
La$_{0.5}$Ca$_{0.5}$MnO$_3$. In fact, systematic studies on
R$_{1-x}$Ba$_x$MnO$_3$ (R= rare-earth) are also scarce, which prompted
Barnabe et al\cite{12} to investigate phase transitions and
magnetoresistance in the R= La series, though there are  some 
reports on specific compositions close to x= 0.3 (Ref. 6, 7, 9, 13-17). It
is also to be noted that the magnetic behavior of La$_{1-x}$Ca$_{x}$MnO$_3$
is extremely sensitive to sample preparative conditions and impurities and
that the magnetic phase diagram is very complex around x= 0.5 [See, for
instance, 8,  18, 19, 20]. Therefore, it is absolutely essential to keep the
starting materials and sample-preparative conditions the same for a series
of compounds, if one has to compare the properties of several compositions
within such a series.  We have therefore taken sufficient care in the
present investigation in this respect.

Polycrystalline samples of the series, La$_{0.5}$Ca$_{0.5-x}$Ba$_x$MnO$_3$
(x= 0.0, 0.1, 0.2, 0.3, 0.4 and 0.5),  have been prepared by a standard
solid state reaction route using required amounts of high purity (better
than 99.9\%) La$_2$O$_3$, CaCO$_3$, BaCO$_3$ and Mn$_2$O$_3$ with heat
treatment similar to that mentioned in Ref. 19, except that the final
sintering was done for 7 days.  We have to necessarily perform final
sintering for such a long duration, as otherwise the ferromagnetic
transitions in these compounds are found to be spread over a wider
temperature range.  The x-ray diffraction patterns taken with
Cu-K$_\alpha$ radiation are shown in Fig. 1; the positions of the
diffraction lines  and the lattice constants derived from these patterns
(table 1)  clearly highlight systematic lattice expansion with the
substitution of Ba for Ca.  The data were analysed by the Rietveld method
using the program FULLPROF.\cite{21} We find that the samples are single
phase for $x<$0.4, forming in an orthorhombic structure with the space
group Pnma (Ref. 22).  For higher compositions there are additional weak
lines which appears to arise from an extra phase, identified as BaMnO$_3$
(also see Ref. 20); it appears that, even for $x$= 0.5, we are able to
retain orthorhombic structure in contrast to the cubic structure reported
by Barnabe et al\cite{12} and this discrepancy is attributed to different
sample-preparative conditions.  In fact, the value of the well-known
"tolerance factor" (t= [r$_A$+r$_O$]/$\surd$2[r$_B$+r$_O$], a measure of
deviation of the ABO$_3$ perovskite structure from cubicity), listed in
table 1 for all compositions, is 0.976, well below unity for this
composition. Therefore, it is not unnatural that this composition is
non-cubic. The $\rho$ measurements were done by a four probe method in the
presence and the absence of a magnetic field (H) employing a conducting
silver-paint for electrical contacts; the magnetoresistance (MR=
$\Delta$$\rho$/$\rho$= [$\rho$(H)-$\rho$(0)]/$\rho(0)$) measurements were
performed in the longitudinal mode.   Dc magnetic susceptibility ($\chi$)
measurements (H= 2000 Oe and 100 Oe; zero-field-cooled (ZFC)) were
performed employing a superconducting quantum interference device (SQUID)
in the temperature range 4.2-350 K; the data for the field-cooled (FC)
state of the specimens were also collected in the presence of H= 100 Oe;
isothermal magnetization (M) measurements were performed at 4.2 K up to 55
kOe. Ac $\chi$ measurements (4.2-350 K), with a frequency of 105 Hz and a
driving field of 0.8 Oe, were also performed both while cooling as well as
warming the samples.

The results of $\rho$ measurements are shown in Figs. 2 and 3. As known in
the literature, $x$= 0.0 compound shows insulating behavior as a result of
which the values of $\rho$ below 50 K are not measurable; however,  the
application of H, say 30 kOe, results in a significant reduction of low
temperature $\rho$ due to the melting of the CO state (Fig. 3a). The same
is qualitatively the case for x= 0.1 as well (see Fig. 3b). Further
substitution of Ba results in gradual reduction of $\rho$ (see Fig. 2),
showing the transformation of the insulating ($x$=0.0) state to a metallic
($x$=0.5) phase.  The intermediate compositions ($x$= 0.2, 0.3, and 0.4)
exhibit negative temperature coefficient of $\rho$ (semi-conducting-like)
below 300 K followed by a broad hump at around 50, 100 and 200 K
respectively; these humps possibly arise from grain boundary effects
[6,7].  Thus, the resistivity values, though large, are in the convenient
range of measurement down to 4.2 K even in zero-field, unlike $x$= 0.0 and
0.1 compounds.  It is to be noted that, in the $x$=0.2 and 0.3 compounds,
there is a weak increase in $\rho$ at low temperatures (T $<$ 30 K, see
Fig. 3c).  Recently, there have been reports [6,7] on A-site mismatch
induced electron-localization in some of the rare earth manganites. The
relevant parameter which characterizes this disorder is the variance
($\sigma^2$) (Ref. 17) of ionic radii at the A-site, and this parameter is
rather large for $x$= 0.2 and 0.3 of Ba as compared to Sr doping (see
table 1). Hence, disorder induced electron-localization effects may be
operative in the Ba-substituted compounds. The application of magnetic
field for these intermediate compositions depresses the magnitude of
$\rho$ at low temperatures and it must be emphasized that the magnitude of
this depression is larger at temperatures far below respective T$_C$
values (see below). Clearly, the magnitude of (negative) MR is large and
precisely measurable in the temperature range 4.2-100 K (-70 to -80$\%$;
see Fig. 3c, inset) for these compositions, unlike the situation in $x$=
0.0 and 0.1.  Following the arguments in Refs. 6 and 7, we attribute the
low temperature enhancement of MR to size-mismatch effects across the
grain boundaries.  When measured as a function of H at 4.2 K, $\rho$
undergoes a sharp drop for initial application of H due to the dominant
scattering response across grain boundaries, followed by a slower fall
with further increase of H (Fig.  3d).  MR exhibits hysteretic behavior
apparently due to such grain boundary effects in ferromagnets.    For the
compositions at the Ba rich end, which are metallic, there is no
worthwhile feature in MR and hence we do not present these results here.

In order to understand the magnetism of these oxides better, we have
performed magnetization studies as well. In the dc $\chi$ vs T data (Fig.
4),  we find a sharp increase of $\chi$ for all the compositions below a
certain temperature, attributable to the onset of ferromagnetic ordering.
The value of T$_C$ inferred from this rise (around 250, 120, 200, 235, 300
and 340 K for x= 0.0, 0.1, 0.2, 0.3, 0.4 and 0.5) undergoes non-monotonous
variation with  Ba substitution.  Even if the heat treatment is the same
as that employed in Ref. 8, we see a similar trend, though these transitions
are very broad for such short-time heat treatments. It may be recalled
that, in the Sr-substituted samples [8], T$_C$ monotonically increases
with $x$, which is consistent with the well-known direct relationship
between A-site ionic size and DE-mediated ferromagnetism. The breakdown of
this relationship in the present series should arise from the strains in
the lattice induced by initial Ba substitution. We therefore interpret the
observed non-monotonic trends in T$_C$ in the Ba substituted compounds in
terms of a competition between the disorder-induced electron-localization
and an enhancement of the DE-induced ferromagnetism resulting from an
increase in the angle subtended by Mn-O-Mn bond and/or a decrease in the
O2p covalent mixing strength with increasing Ba concentration.\cite{10}
Similar non-monotonic variation of T$_C$ due to size-mismatch effects has
been known, to our knowledge, in only one system, viz.,
Pr$_{0.5}$(Ba,Sr)$_{0.5}$MnO$_3$ (Ref.  11) and, in this sense, our
finding gains importance.

We note the following minor discrepancies in the observed values of T$_C$
with those reported in the literature for the two end members. It is well
known [8, 19] that the $x$= 0 compound orders ferromagnetically at 225 K;
slightly higher T$_C$ value for this composition in the present studies
must be due to different heat-treatment conditions, as the starting raw
materials are the same as those employed in Ref. 8. In the same way, our
T$_C$ value for x= 0.5 is larger than that reported (about 290 K) by
Barnabe et al,\cite{12} which could be attributed to cubicity of our
sample as mentioned earlier. It may be added that, on the basis of
investigations on La$_{0.67}$Ba$_{0.33}$MnO$_3$, Ju et al\cite{6} conclude
that oxygen deficiency lowers the value of T$_C$ and hence we believe that
the oxygen non-stoichiometry (deviation from 3) is relatively less in our
samples.   For $x$= 0.0, we note that the drop in $\chi$ due to the onset
of antiferromagnetism below 150 K (Ref. 8 and 19) is virtually absent,
though the reduced magnitude of high-field magnetization (see below) is
indicative of the existence of an antiferromagnetic component; presumably,
the present sample preparative conditions induce canting in the spin
structure inducing a ferromagnetic component for this composition even
below about 150 K; this low temperature ferromagnetic component cannot be
of DE-mediated type unlike the one setting in around 250 K, as otherwise
one should have observed metallicity at low temperatures.

\vskip 0.6cm
There are also other interesting features in the dc and ac magnetization
data: 
\begin{itemize}
\item
As is well-known, ac $\chi$ shows (Fig. 5) hysteretic behavior while
cooling and warming in the temperature range 75-140 K for $x$= 0.0 due to
first-order nature of the charge-ordered antiferromagnetic
transition.\cite{8} Interestingly, the same hysteretic behavior is seen
for $x$= 0.1 as well around the temperature at which there is an onset of
{\it ferromagnetic} ordering. It is not clear whether this signals the
persistence of CO around 100 K for this composition as well (nearly at the
same temperature as T$_C$!).  In any case, our data seem to
provide evidence for the first-order nature of the ferromagnetic
transition similar to the observation on
La$_{0.52}$Gd$_{0.15}$Ca$_{0.33}$MnO$_3$ (Ref. 23), a behavior not so
common among the family of   Mn-based perovskites.

\item
It is  interesting to note that that the composition $x$= 0.1 is
insulating and hence this  sample is not a DE-mediated ferromagnet.
Thus this series provides an opportunity to traverse from spin-canted,
charge-ordered, insulating state to a  metallic, DE-ferromagnet via a
(charge-ordered?) insulating ferromagnet by varying Ca/Ba ratio.

\item
The isothermal magnetization (M) (Fig.  6) rises sharply for initial
applications of H (below 10 kOe) for all compositions. M saturates to a
value close to 3.1 $\mu$$_B$ for all compositions, except for $x$= 0.0,
typical of that known for the ferromagnets of this class of
Mn-perovskites; the values are however low (close to 1 $\mu$$_B$)  at high
fields for $x$= 0.0, without saturation even at high fields; this
observation attests our earlier remark that this alloy is essentially a
canted antiferromagnet for the present sample preparative conditions.  It
is interesting to note that in the Sr doped compounds,\cite{8} the low
temperature antiferromagnetic state is enhanced for initial Sr
substitution (up to $x$= 0.2), whereas for Ba doping, $x$= 0.1 is
sufficient to depress antiferromagnetism. If an antiferromagnetic
interaction coexists with ferromagnetism for $x$ (Ba)= 0.1, it may occur
below 50 K as evidenced by a shoulder in ZFC dc and ac magnetization (see
Fig. 4 and 5). However this component is suppressed/modified by the
presence of H or field cycling. Further observations made below render
additional support to this view and the canted nature of low temperature
magnetic structure.

\item
In the M versus H plots (Fig.  6), for x= 0.0 and 0.1, as the field
is increased, there is a strong irreversibility between the  initial
magnetization while increasing the field and the one while decreasing the
field, as shown in Fig. 6. These features are similar to those reported in
Nd$_{1-x}$Ca$_x$MnO$_3$ (Ref.  24).   For x= 0.1, we observe a more
spectacular feature in the sense that, in the initial curve, there is a
sudden increase in M at about 35 kOe. We attribute these features to
possible small canting of magnetic structures, which is presumably altered
by the application of H.

\item
The coercive field values are found to be typically 100 Oe for $x$$<$
0.3, beyond which the value is still closer to zero; these values suggest
that the samples are soft ferromagnets. (vi) The ZFC and FC (H= 100 Oe) dc
$\chi$ curves tend to deviate from each other from a lower temperature
compared to respective T$_C$ values,  presumably due to the anisotropy of
the materials (plots shown only for three compositions in Fig. 4b for the
sake of clarity of the figure). Similar divergencies have been known even
in other magnetic materials exhibiting long range magnetic ordering and
hence need not be associated with spin-glass ordering.\cite{25} The
percentage of the divergence ultimately at low temperatures, however,
decreases with the increase of Ba concentration, which we attribute to
corresponding decrease of metastability.

\end{itemize}

Thus, we have traced the transformation of the CO insulating state ($x$=
0) into a ferromagnetic-like metal ($x$=0.5) with increasing Ba
concentration in La$_{0.5}$Ca$_{0.5-x}$Ba$_x$MnO$_3$, interestingly
traversing through an insulating ferromagnetic compound with a
non-monotonic variation of T$_C$ with x. We made several interesting
observations in the magnetic and transport behavior of these oxides.  For
$x$= 0.2 and 0.3, the magnetoresistance is quite large and precisely
measurable over a wide temperature range well below T$_C$, an observation
which may be of importance for applications.  We would like to add that,
at the time of writing this article, we came across a publication by Zhong
et al\cite{26} reporting the observation of a large MR over a wide
temperature range in another manganite,
La$_{0.55}$Dy$_{0.12}$Ca$_{0.33}$MnO$_3$, and thus there is a considerable
interest in identifying systems with such properties.  The results
establish that the A-site ionic-size mismatches play a decisive role on
the transport and magnetic behavior in these manganites.

\rightline{$^*$Electronic address: sampath@tifr.res.in}

\begin{figure}
\caption{X-ray diffraction patterns  at 300 K of some
compositions of the series, La$_{0.5}$Ca$_{0.5-x}$Ba$_x$MnO$_3$. The
asterisks for Ba rich compositions mark the lines arising from an extra
phase, presumably due to BaMnO$_3$. The pattern around the most intense
lines are shown in an expanded form in the inset.}
\end{figure}

\begin{figure}
\caption{Electrical resistivity ($\rho$) as a function of temperature for
the compounds, La$_{0.5}$Ca$_{0.5-x}$Ba$_x$MnO$_3$.}
\end{figure}

\begin{figure}
\caption{ Temperature dependent electrical resistivity ($\rho$) behavior
for $x$= 0.0, 0.1, 0.2 and 0.3 samples of the series,
La$_{0.5}$Ca$_{0.5-x}$Ba$_x$MnO$_3$, in zero magnetic field and in a field
of 30 kOe is shown in (a), (b) and (c). The temperature dependent
magnetoresistance for $x$= 0.2 and 0.3 is shown in the inset of (c).
$\rho$ as a function of magnetic field (H) at 4.2 K for $x$= 0.2 and 0.3
is shown in (d).}
\end{figure}

\begin{figure}
\caption{(a) Dc magnetic susceptibility ($\chi$) as a
function of temperature measured in a magnetic field of 2 kOe and (b)
typical zero-field-cooled and field-cooled $\chi$ behavior measured in a
field of 100 Oe, for the compounds, La$_{0.5}$Ca$_{0.5-x}$Ba$_x$MnO$_3$.}
\end{figure}

\begin{figure}
\caption{Temperature dependent ac susceptibility behavior, for the
compounds, La$_{0.5}$Ca$_{0.5-x}$Ba$_x$MnO$_3$. For two compositions, $x$=
0.0 and 0.1, we observe hysteresis behavior while cooling and warmimg
(shown separately in the upper part of the figure).}
\end{figure}

\begin{figure}
\caption{Isothermal magnetization behavior at 4.2 K for the series,
La$_{0.5}$Ca$_{0.5-x}$Ba$_x$MnO$_3$. The data for $x$= 0.4 is practically
the same as that for $x$= 0.5.  For x= 0.0 and 0.1, the deviation of the
field-cycled magnetization from the initial magnetization (that is, for
increasing magnetic field) is shown in the insets.}
\end{figure}

\begin{table}
\caption{Lattice parameters, average A-site ionic radius
($<$r$_A$$>$), tolerance factor (t), variance ($\sigma$$^2$) of A-site
ionic size and Curie temperature (T$_C$) in the
La$_{0.5}$Ca$_{0.5-x}$Ba$_x$MnO$_3$ compounds. The $\sigma$$^2$ values for
the Sr compounds are also given. The errors in the lattice parameters are
estimated to be 0.004 $\AA$.}
\begin{tabular}{l|ccccccc}

$x$ & a $(\AA)$ & b $(\AA)$ & c $(\AA)$ & $<$r$_A$$>$ & t & 
        $\sigma$$^2$ Ba(Sr)  & T$_C$ (K) \\ \hline
0.0 & 5.413   & 7.631   & 5.424   & 1.198 & 0.923 & 0.0003 (0.0003) & 250\\
0.1 & 5.437   & 7.655   & 5.461   & 1.227 & 0.933 & 0.0069 (0.0014) & 120\\
0.2 & 5.487   & 7.713   & 5.501   & 1.256 & 0.944 & 0.0117 (0.0021) & 200\\
0.3 & 5.480   & 7.731   & 5.509   & 1.270 & 0.954 & 0.0148 (0.0025) & 235\\
0.4 & 5.499   & 7.759   & 5.511   & 1.285 & 0.965 & 0.0163 (0.0025) & 300\\
0.5 & 5.522   & 7.793   & 5.533   & 1.314 & 0.976 & 0.0162 (0.0022) & 340
\end{tabular}
\end{table}

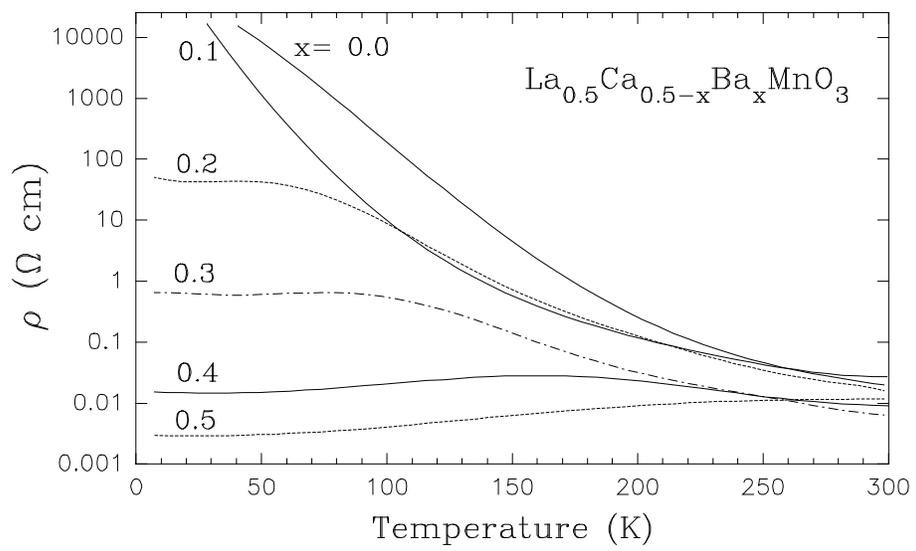
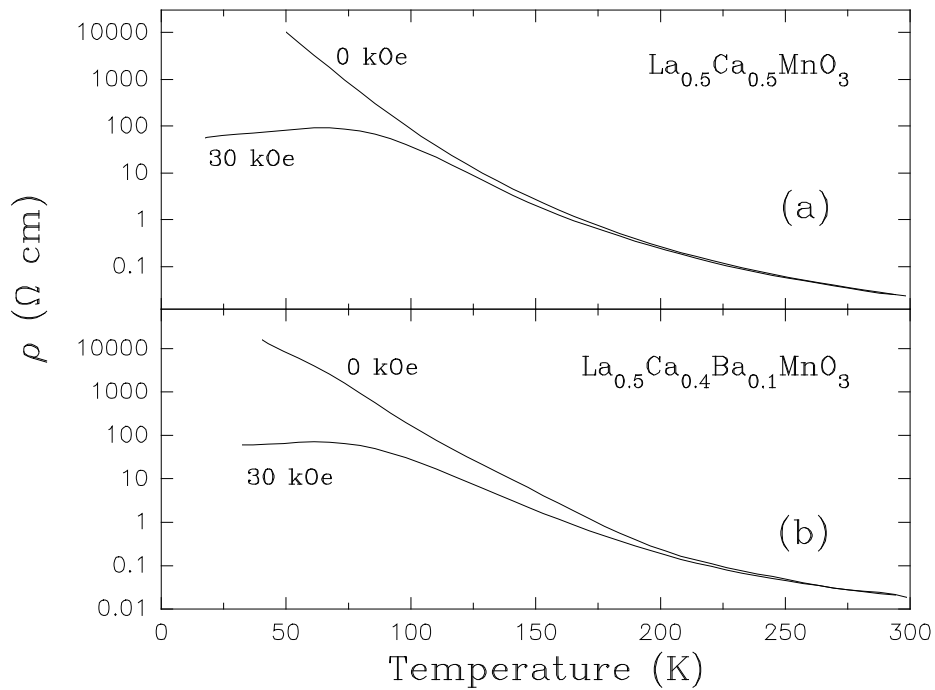
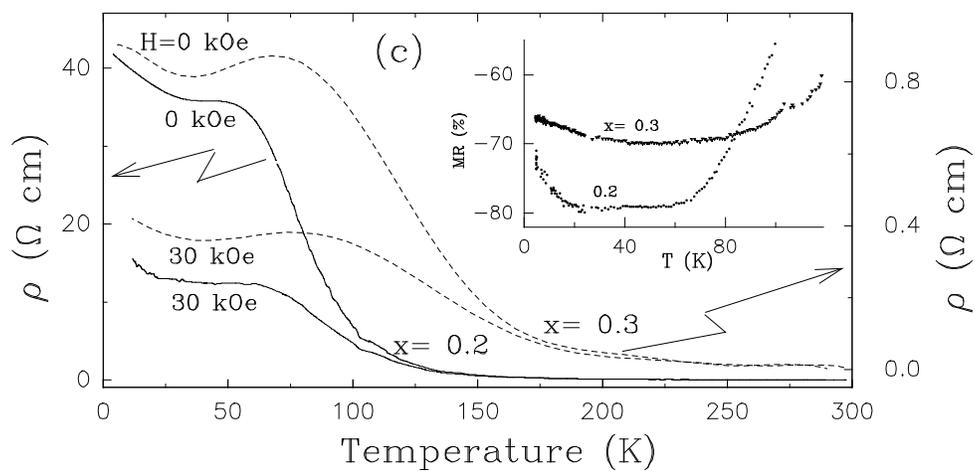
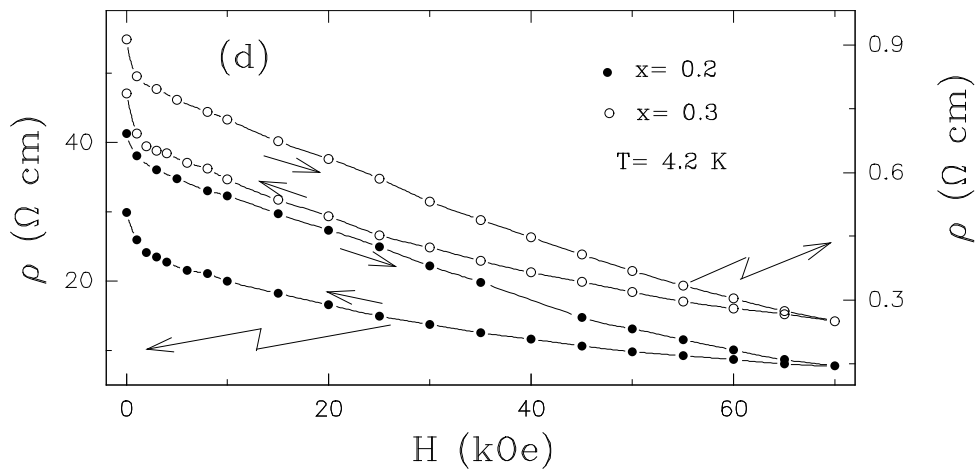
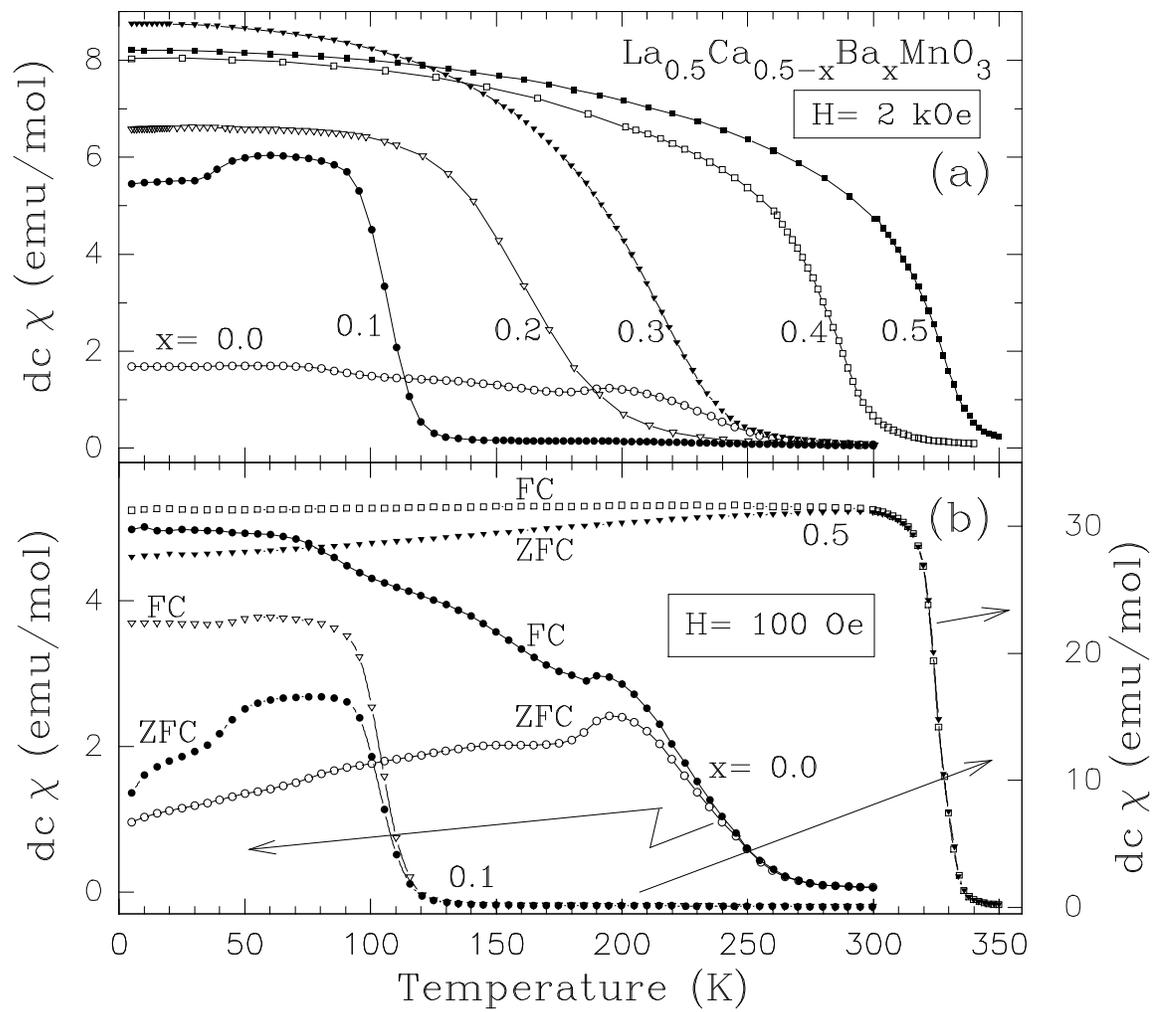
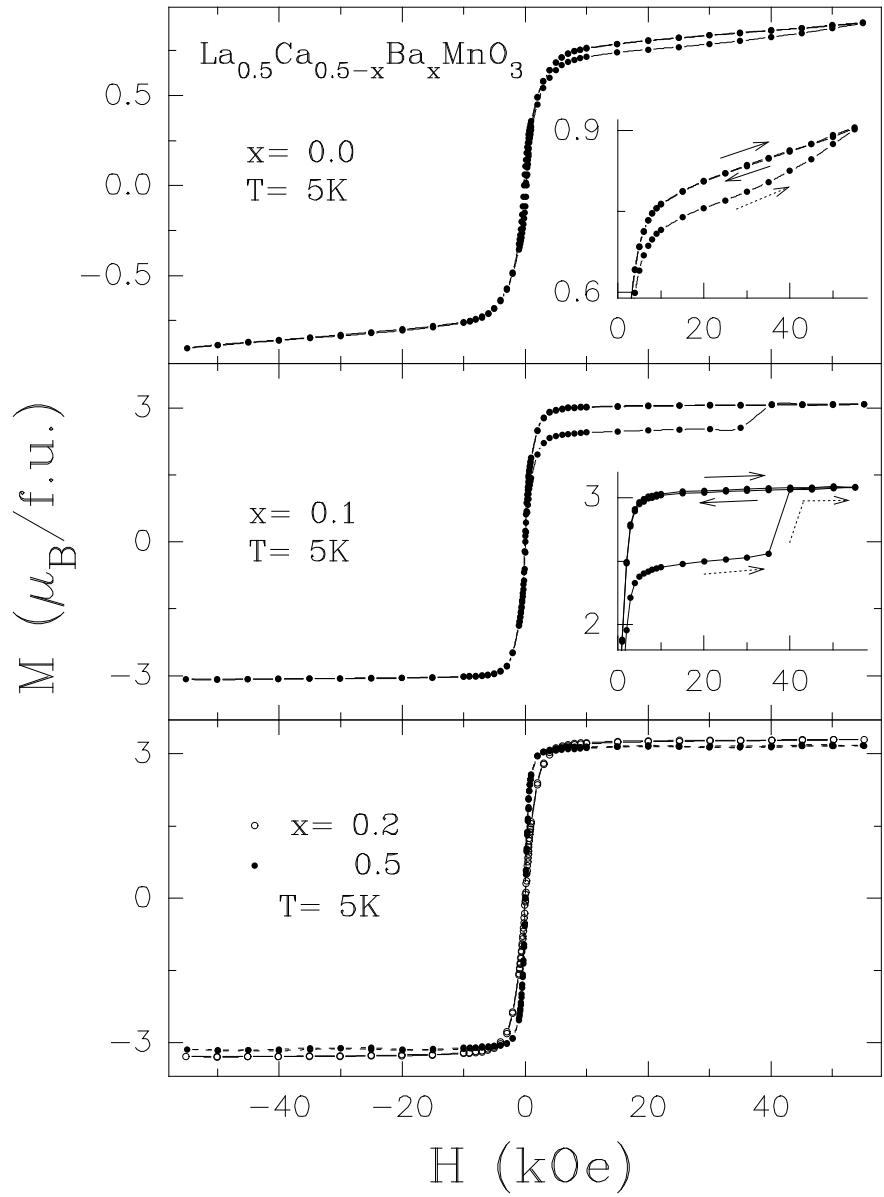
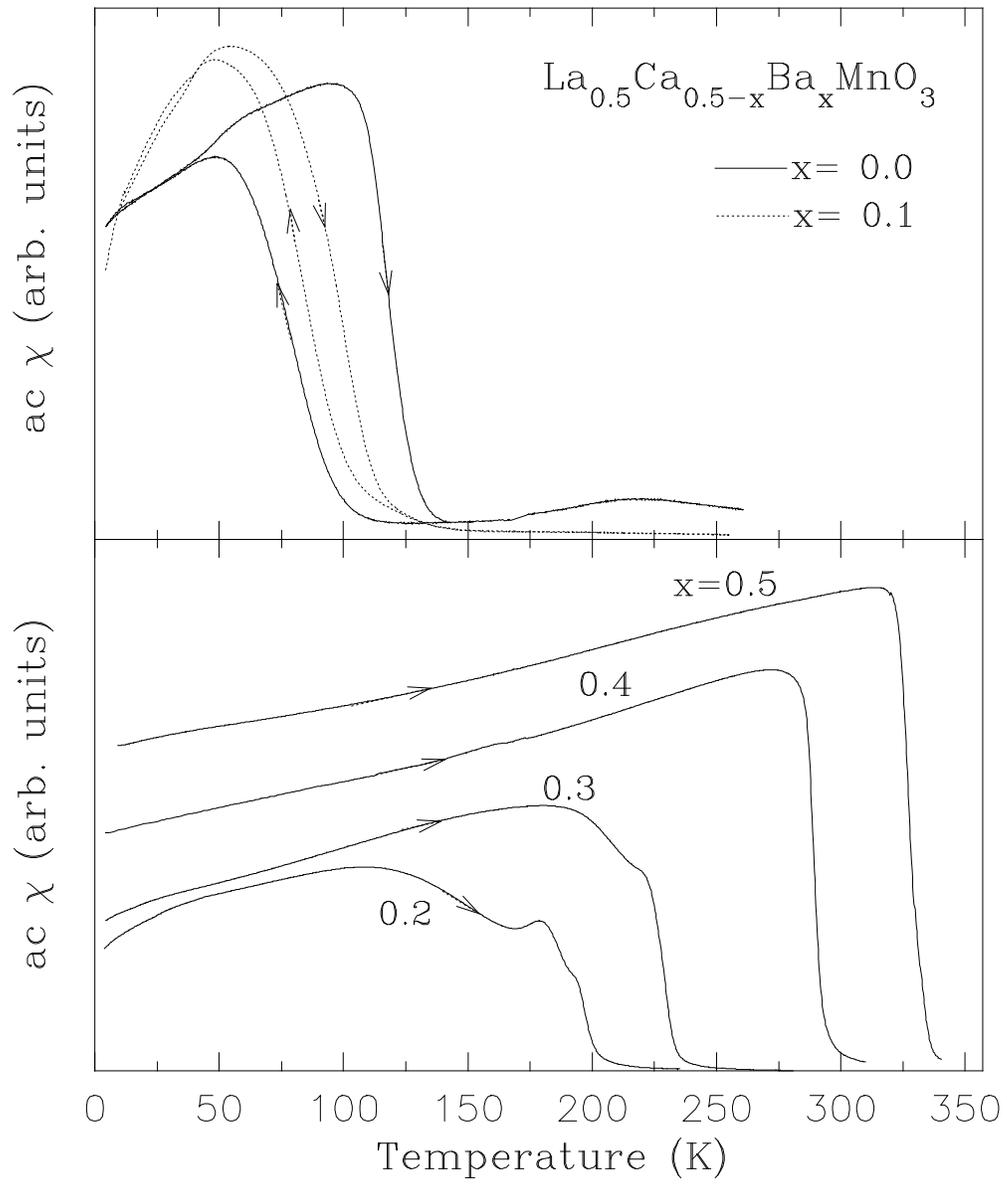
\end{document}